\documentclass[aps,pra,twocolumn,showpacs]{revtex4}

\usepackage{times,amsmath,amssymb,graphicx}
\begin{document}
\renewcommand{\vec}[1]{\mbox{\boldmath$\mathrm{#1}$}}
\newcommand{\ddp}[3]{\frac{\partial^2 #1}{\partial #2\partial #3}}
\newcommand{\dddp}[4]{\frac{\partial^3 #1}{\partial #2\partial #3\partial #4}}

\title{Dynamic hyperpolarizability of the one-dimensional hydrogen
  atom with a $\delta$-function interaction}


\author{K. Satitkovitchai}
\affiliation{Theoretical Chemistry, Zernike Institute for Advanced Materials, University of Groningen,
Nijenborgh 4, 9747 AG Groningen, The Netherlands}
\date{\today}

\begin{abstract} 
 The dynamic hyperpolarizability of a particle bound by the one-dimensional
  $\delta$-function potential is obtained in closed form. On the first step, we 
  analyze the singular structure of the non-linear response function as
  given by the sum-over-state expression. We express its poles and residues in
  terms of the wave-number $k$.  On the second step, we calculated the
  frequency dependence of the response function by integration over
  $k$. Our method provides a unique opportunity to check the
  convergence of numerical methods, and is in a perfect agreement with the 
  static and high frequency limits obtained by different theories.
  The former is obtained using the approach of Swenson and Danforth
  (J. Chem. Phys. {\bf 57}, 1734 (1972)). The asymptotic decay is studied using
  the method of Scandolo and Bassani (Phys. Rev. B {\bf 51}, 6925 (1995)). Its
  extension to the case of quadrupole polarizability reveals a universal (not
  dependent on the choice of the system) asymptotic behavior of the
  hyperpolarizability. 
\end{abstract}


\maketitle 
\section{Introduction}
There is only a small number of exactly solvable realistic quantum
systems. The hydrogen atom is one of the most experimentally and theoretically
studied. Remarkably, expressions for the Green function~\cite{Hostler1964},
transition matrix elements between the ground and excited
states~\cite{Bethe1957}, and a number of other properties of this system exist
in closed form.  However, the complexity of the matrix elements as well as of
the sum-over-states (SOS) expressions for the nonlinear optical response
functions~\cite{Orr} hinders an analytic computation of its dynamic
hyperpolarizabilities. Already expressions for the dynamic polarizability of
hydrogen atom are very complicated~\cite{Au1978}.

To describe the multiphoton ionization of atomic hydrogen it is sufficient to
consider summations over all intermediate bound states~\cite{Bebb1966}. In
contrast, the computation of the nonlinear optical response is much more
involved since summations over the continuum states should also be
performed~\cite{Tang1986}. This considerably complicates derivations and calls
for the development of new methods. Thus, Mizuno~\cite{Mizuno1972} used the
sturmian coulomb Green function~\cite{Maquet1977} in order to compute the
third harmonic generation coefficient of the hydrogen atom. These results were
extended by Shelton~\cite{Shelton1987} to other third order
processes. Although written in analytical form, they still contain infinite
sums, and thus can only be analyzed numerically. Therefore, it is desirable to
have a simpler model system that would allow for a closed-form solution for
non-linear optical responses.

A particle bound by a one-dimensional $\delta$-function potential bares a
close resemblance to the three-dimensional hydrogen atom with Coulomb
potential. Despite its simplicity, the system is attractive for the
theoretical analysis because like the hydrogen atom it contains both discrete
and continuum eigenstates. Also the Green function of these two systems shows
remarkable analogies~\cite{Blinder1988}. The $\delta$-potential allows for a
great simplification of the analytic work since one only needs to carry out
the integration of relatively simple functions. 

 The frequency-dependent electric polarizability of this system has previously
been computed by explicit solution of the time-dependent Schr\"{o}dinger
equation with uniform time-dependent electric field and expansion of the
charge density in the perturbed ground state up to and including terms linear
in applied electric fields~\cite{Postma1984}. Recently, the result has been
confirmed using a perturbation technique~\cite{Maize2004} based on the work by
Nozi\'{e}res~\cite{Nozieres}. Here, we present results for the \emph{lowest}
order non-zero hyperpolarizability due to the combination of two
electric-dipole and one electric-quadrupole transitions. This response
function describes the second-harmonic generation (SHG) process. SHG due to
only dipole transitions is forbidden because the system possesses an inversion
symmetry. The next order non-linear response (the third harmonic generation)
due to the combination of four dipole transition should have a comparable magnitude
and will be analyzed elsewhere.

As a starting point, in Sec.~II we study singular properties of the SHG
response function as given by equations of Orr and Ward~\cite{Orr}. We assume
that all conditions of validity of these formulas are fulfilled. The reader is
referred to the book of Shen~\cite{Shen} for the discussion of relevance of
these expressions for a certain experimental situation. Using expressions for
the poles and residues of the response function obtained here we perform the
actual computation for the one-dimensional hydrogen in
Sec~\ref{sec:main}. Finally, we qualitatively analyze obtained results in the
low and high frequency limits and compare them with prediction of other
theories (Sec.~\ref{sec:limits}).

In this respect we refer to the work of Fern\'{a}ndez and
Castro~\cite{Fernandez1985} which gives an introduction to the method that
avoids an integration over the continuum spectrum. Based on the original idea
of Swenson and Danforth~\cite{Swenson1972}, and extending the derivation of
Austin~\cite{Austin1979}, they obtained a recursive equation that relates the
expectation values of $x^N$ operator at different level of the perturbation
expansion. They used the method to find a perturbative correction to the ground
state energy and to describe the appearance of resonant states in the presence
of static electric field. As will be shown below, the same approach can be
used to obtain an arbitrary order static hyperpolarizability~\cite{remark}. 
We performed corresponding calculations for our model system and found an 
agreement with the static limit of our SOS-derived expression.

The high frequency limit of the second-harmonic generation in dipole
approximation was obtained by Scandolo and Bassani~\cite{Scandolo1995}. Later,
this result was generalized to higher harmonic generation processes using
the general quantum theory of Kubo optical response
function~\cite{Bassani2000}. It yields $1/\omega^{2n+2}$ decay of $n$-th order
response function. We show that non-local response (the quadrupole
polarization) dictates a different asymptotic behavior, $1/\omega^4$ for the
SHG process, in agreement with our SOS results.

 In Appendix~\ref{sec:apa}, we outline calculations of the transition matrix
elements for this system, and in Appendix~\ref{sec:apb} we derive the
frequency dependence of the SHG response function from the corresponding
residues. An extensive use of the complex analysis and generalized functions
is made. All calculations are done analytically. The final results are
expressed in terms of elementary functions.
\section{Singularities of the SHG response function}
The poles and residues of the second-order response functions can be derived
from the microscopic expressions for the nonlinear polarizations (as, for
example, given by Eq.~(43) of Ref.~\onlinecite{Orr}) assuming certain form of
the light-matter interaction. It is important to notice that corresponding
perturbation operators $V$ as well the polarization operator $\vec P$ must be
used in the renormalized form with vanishing expectation values in the ground
state
\begin{eqnarray}
V&=&\tilde{V}-\langle \tilde{V}\rangle_{gg},\label{eq:pert}\\
\vec P&=&\tilde{\vec P}-\langle \tilde{\vec P}\rangle_{gg},\label{eq:polariz}
\end{eqnarray}
where $ \tilde{V}$ and $\tilde{\vec P}$ are the bare operators.  Other forms
of presenting the non-linear response functions are also known. For example,
lifting the requirement on the diagonal elements to be zero leads to the
appearance of additional terms~\cite{Langhoff1972}.  They can be traced back
to the secular divergence problem. Clearly, both forms lead to
identical results at the end. To make our manuscript self-contained, we reproduce here
the equation of Orr and Ward for a general second-order process
($\omega_\sigma=\omega_1+\omega_2$) using slightly simplified notations:
\begin{widetext}
\begin{eqnarray}
\vec P^{\omega_\sigma}&=&\frac{K(\omega_1,\omega_2)}{(-\hslash)^2}I_{1,2}
\sum\limits_{m,n}\Biggl\{
\frac{\langle \vec P\rangle_{gm}\langle V^{\omega_2}\rangle_{mn}\langle V^{\omega_1}\rangle_{ng}}
     {(\Omega_{mg}-\omega_\sigma)(\Omega_{ng}-\omega_1)}
+\frac{\langle V^{\omega_2}\rangle_{gm}\langle V^{\omega_1}\rangle_{mn}\langle\vec P\rangle_{ng}}
     {(\Omega^*_{mg}+\omega_2)(\Omega^*_{ng}+\omega_\sigma)}
+\frac{\langle V^{\omega_2}\rangle_{gm}\langle\vec P\rangle_{mn}\langle V^{\omega_1}\rangle_{ng}}
     {(\Omega^*_{mg}+\omega_2)(\Omega_{ng}-\omega_1)}
\Biggr\}.\label{eq:p2}
\end{eqnarray}
\end{widetext}
The simplification concerns the use of only renormalized operators, while in
the original formulation the authors had both renormalized and bare
operators. Such transformation is always possible since $\langle
V\rangle_{ng}=\langle \tilde{V}\rangle_{ng}$ (the same also holds for $\vec
P$). As a consequence, the expression acquires a more symmetrical form.  Here,
$I_{1,2}$ denotes the average of all terms generated by permuting $\omega_1$
and $\omega_2$. $K$ is a numerical factor that depends on the permutational
symmetry of incident photons. For the SHG process it is equal to
$\frac12$. The summations are running over all excited states excluding the
ground state.  $\Omega_{mg}$ denotes the energy difference between the excited
state $m$ and the ground state (labeled by $g$). In order to avoid
divergencies at the resonances, a phenomenological damping ($i\Gamma$) of
excited states is introduced, $\Omega_{mg}=E_m-E_g+i\Gamma$. This shifts the
poles of the response functions away from the real axis on the complex plane.

If the damping constant $i\Gamma$ is small, it has no relevance for the
derivation of residues. That is why we provisionally assume that all poles lie
on the real axis ($\Omega^*_{ng}=\Omega_{ng}$). Furthermore, a simple analysis
of Eq.~(\ref{eq:p2}) shows that the poles of nonlinear polarization are
symmetrically situated around $\omega=0$. Additional symmetry also exists for
their residues. The residue of a pole at negative energy has opposite sign to
the residue of its positive energy counterpart. Thus, for our analysis it is
sufficient to consider singularities at positive energies only.

A simple analysis of the SHG case ($\omega_1=\omega_2=\omega$,
$\omega_\sigma=2\omega$) shows that each excited state $m$ gives origin to 4
poles on the complex plane $\omega=\pm\Omega_{mg}$ and $\omega=\pm
\Omega_{mg}/2$. The SHG response function is defined as a second derivative
with respect to the external electric field
\begin{equation}
\chi^{(\omega,\omega)}_{i;jk}
   =\frac12\ddp{P^{2\omega}_i}{F^{\omega}_j}{F^{\omega}_k}.\label{eq:chi-def}
\end{equation}
Its residues for positive frequencies will be denoted as follows:
\begin{subequations}
\label{eq:AB-shg}
\begin{eqnarray}
A^{(\omega,\omega)}_{i;jk}(m)&=&\mathrm{Res}_{\omega=\Omega_{mg}}\chi^{(\omega,\omega)}_{i;jk},
\label{eq:A-shg}\\
B^{(\omega,\omega)}_{i;jk}(m)&=&\mathrm{Res}_{\omega=1/2\Omega_{mg}}\chi^{(\omega,\omega)}_
{i;jk}.\label{eq:B-shg}
\end{eqnarray}
\end{subequations}
There will be two terms contributing to the residue of the polarization at
$\omega=\Omega_{ng}$:
\begin{widetext}
\begin{equation}
\mathrm{Res}_{\Omega_{ng}}\vec
P^{2\omega}=\frac{-1/2}{(-\hslash)^2}\sum\limits_{m}
\Biggl\{
\frac{\langle \vec P\rangle_{gm}\overline{\langle V^{\omega}\rangle_{mn}\langle V^{\omega}\rangle_
{ng}}}
     {\Omega_{mg}-2\Omega_{ng}}
+\frac{\langle\vec P\rangle_{mn}\overline{\langle V^{\omega}\rangle_{ng}\langle V^{\omega}\rangle_
{gm}}}
     {\Omega_{mg}+\Omega_{ng}}
\Biggr\}.\label{eq:p2-shg-a}
\end{equation}
\end{widetext}
Here, we substituted the frequencies and the symmetry factor $K$ and expanded
the symmetrization operator. A bar over the matrix elements denotes their
symmetrization, which signifies the equivalence of two incident photons and
follows from the application of $I_{1,2}$ to the whole expression. In the same
way, we write an expression for the residue of the polarization at
$\omega=\Omega_{mg}/2$ where only one term contributes:
\begin{equation}
\mathrm{Res}_{\Omega_{mg}/2}\vec
P^{2\omega}=\frac{-1/4}{(-\hslash)^2}\sum\limits_{n}
\Biggl\{
\frac{\langle \vec P\rangle_{gm}\overline{\langle V^{\omega}\rangle_{mn}\langle V^{\omega}\rangle_
{ng}}}
     {\Omega_{ng}-1/2\Omega_{mg}}\label{eq:p2-shg-b}
\Biggr\}.
\end{equation}

Equations (\ref{eq:p2-shg-a}) and (\ref{eq:p2-shg-b}) are obtained from the
general theory, and, therefore, are valid for systems of arbitrary
dimensionality and for different light-matter interaction mechanisms. They are
not only useful in present theoretical analysis, but also can bring a
substantial computational savings when used in \emph{ab initio} calculations
for realistic systems. This directly follows from the estimates on the number
of floating-point operations needed to directly evaluate Eq.~\ref{eq:p2}
($N_{sos}=O(N_\omega\cdot N^2$)) in comparison with a two-step procedure,
where the residues (Eqs.~\ref{eq:p2-shg-a},\ref{eq:p2-shg-b}) are computed
with numerical cost of $N_{sos}^I=O(N^2)$ only, and the frequency dependence
is obtained on the second step using $N_{sos}^{II}=O(N_\omega\cdot N)$
operations. Here we assumed that the system has $N$ excited states and
$N_\omega$ frequency points are required.

On the last step, we specify Eqs.~(\ref{eq:p2-shg-a},\ref{eq:p2-shg-b}) for
the model system. Since the SHG is strictly forbidden in centrosymmetric
systems within the electric dipole approximation
($\chi_{eee}^{(\omega,\omega)}=0$), we will be considering a one-component
response function $\chi_{qee}^{\omega,\omega}$ with the dipole perturbation
operator $\tilde{V}=-exF$, and the quadrupole polarization
$\tilde{P}=ex^2$. Furthermore, we make use of atomic units, the conversion is
done by setting the electron charge $e=-1$, and $m=\hslash=1$. Finally, the
differentiations with respect to the external electric field $F$ are performed
according to the definition of the response function (Eq.~\ref{eq:chi-def}).

We will denote two contributions to $A_{i;jk}^{(\omega,\omega)}(m)$ as
$A^{I}(k)$ and $A^{II}(k)$, and $B_{i;jk}^{(\omega,\omega)}(m)$ will be
denoted as $B(k)$. It is natural to name them as \emph{spectral functions of
the second-order nonlinear response} in analogy with the spectral functions in
the many-body perturbation theory.
\begin{subequations}
\label{eq:AB-shg-res}
\begin{eqnarray}
A^{I}(k)&=&-\frac12\sum\limits_{q}
\frac{\langle x^2\rangle_{0q}\langle x\rangle_{qk}\langle x\rangle_{k0}}
{\Omega_{q0}-2\Omega_{k0}},\\
A^{I}(k)&=&-\frac12\sum\limits_{q}
\frac{\langle\Delta x^2\rangle_{qk}\langle x\rangle_{k0}\langle x \rangle_{0q}}
{\Omega_{k0}+\Omega_{q0}},\\
B(k)&=&-\frac12\sum\limits_{q}
\frac{\langle x^2 \rangle_{0k}\langle x\rangle_{kq}\langle x \rangle_{q0}}
{2\Omega_{q0}-\Omega_{k0}}.
\end{eqnarray}
\end{subequations}
Instead of a discrete state number $m$, these functions now depend on the
wave-number $k>0$ that characterizes excited states of the system as will be
shown below. In order to be consistent with naming of states of the
one-dimensional hydrogen that will be introduced below we changed the notation
for the ground state to $0$. We also introduced the notation $\Delta
x^2=x^2-\langle x^2 \rangle_{00}$ and took into account that $\langle x
\rangle_{00}=0$ for our model system.
\section{Dynamic hyperpolarizability of the one-dimensional hydrogen\label{sec:main}}
The system discussed here is described by the one-dimensional time-independent
Schr\"{o}dinger equation that can be written in the dimensionless form:
\begin{equation}
H_0\phi=-\phi''/2-\delta(x)\phi=E\phi.\label{eq:Schr}
\end{equation}

The solution of the eigenvalue problem (Eq.~\ref{eq:Schr}) yields a single
bound state with the energy and wave-function:
\begin{equation}
E_0=-1/2,\hspace{1em}\phi_0(x)=\exp(-|x|),
\end{equation}
and a continuum of unbound states~\cite{Postma1984,Lapidus1981}. Since the
Hamiltonian of the system is invariant with respect of space inversion, the
wave-functions can also be constructed to have a well defined parity:
\begin{eqnarray}
\phi_+(k;x)&=&\frac{k}{\sqrt{\pi}\sqrt{1+k^2}}\left[\cos(kx)-\frac1k\sin(k|x|)\right],\\
\phi_-(k,x)&=&\frac1{\sqrt{\pi}}\sin(kx).
\end{eqnarray}
They are degenerate and have the energy $E(k)=k^2/2$ in resemblance with the
unperturbed states of a free particle. However, in contrast to the
free-particle case the wave-number $k$ can only attain \emph{positive} real
values. One can readily demonstrate the completeness and the normalization of
the above set of eigenfunctions.

 As a starting point for $A^{I}(k)$, $A^{II}(k)$ and $B(k)$ computation, we
need the bound-free and free-free transition matrix elements of the electric
dipole:
\begin{widetext}
\begin{subequations}
\begin{eqnarray}
\langle\phi_0|x|\phi_-(k)\rangle&=&\frac4{\sqrt{\pi}}\frac{k}{(k^2+1)^2},\\
\langle\phi_+(k)|x|\phi_-(k')\rangle&=&-\frac{k}{\sqrt{1+k^2}}
\left[\delta^{(1)}(k'-k)-\frac4{\pi}\frac{k'}{(k'^2-k^2)^2}\right],\\
\langle\phi_+(k')|x|\phi_-(k)\rangle&=&\frac{k'}{\sqrt{1+k'^2}}
\left[\delta^{(1)}(k'-k)+\frac4{\pi}\frac{k}{(k'^2-k^2)^2}\right].
\end{eqnarray}
\end{subequations}
For the non-local SHG, we also need the quadrupole transition matrix elements
 between the ground and even unbound states:
\begin{subequations}
\begin{equation}
\langle\phi_0|x^2|\phi_+(k)\rangle=-\frac{8}{\sqrt{\pi}}\frac{k}{\sqrt{1+k^2}}\frac{1}{(1+k^2)^2}
\end{equation}
and between two free odd states:
\begin{equation}
\langle\phi_-(k')|\Delta x^2|\phi_-(k)\rangle=
\langle\phi_-(k')|x^2-\langle\phi_0|x^2|\phi_0\rangle|\phi_-(k)\rangle=
-\left[\frac{d^2}{dk'^2}+\frac12\right]\delta(k'-k).
\end{equation}
\end{subequations}
 The derivation of the matrix elements is done in Appendix~\ref{sec:apa}.

The spectral functions are computed according to
Eqs.~(\ref{eq:AB-shg-res}), replacing the summations with
integrations over the wave-number ($k'$):
\begin{eqnarray}
A^{I}(k)&=&-\frac12\int\limits_{0}^{\infty} 
\frac{\langle\phi_0|x^2|\phi_+(k')\rangle\langle\phi_+(k')|x|\phi_-(k)\rangle\langle\phi_-(k)|x|\phi_0\rangle}
{\omega_0(k')-2\omega_0(k)}dk'\nonumber\\
&=&-\frac{32}{\pi}\frac{k}{(k^2+1)^2}\int\limits_{0}^{\infty}
\frac{1}{2k^2-k'^2+1}\frac{k'^2}{(1+k'^2)^3}
\left[\frac{\delta(k'-k)}{dk'}+\frac1{\pi}\frac{4k}{(k^2-k'^2)^2}\right]dk'
=-\frac{32}{\pi}\frac{k}{(k^2+1)^2}\left[\lambda_1(k)+\frac{4}{\pi}\lambda_2(k)\right]\nonumber\\
&=&-\frac{32}{\pi}\frac{k}{(k^2+1)^2}\left[\frac{2k(k^2-1)}{(k^2+1)^5}+\frac{4}{\pi}\frac{\pi}{32}
\frac{k^4+12k^2-23}{(k^2+1)^5}\right]
=-\frac{4}{\pi}\frac{k^2(k^4+28k^2-39)}{(k^2+1)^7},\label{eq:A1k}
\end{eqnarray}
where we introduced notation $\omega_0(k)=E(k)-E_0=(k^2+1)/2$. Integrals $\lambda_1(k)$
and $\lambda_2(k)$ are computed in Appendix~\ref{sec:apa}.
\begin{eqnarray}
A^{II}(k)&=&-\frac12\int\limits_{0}^{\infty} 
\frac{\langle\phi_-(k')|\Delta x^2|\phi_-(k)\rangle\langle\phi_-(k)|x|\phi_0\rangle\langle\phi_0|x|\phi_-(k')\rangle}
{\omega_0(k)+\omega_0(k')}dk'\nonumber\\
&=&\frac{16}{\pi}\frac{k}{(k^2+1)^2}\int\limits_{0}^{\infty}\left[\frac{d^2}{dk'^2}+\frac12\right]\delta(k'-k)
\frac{k'}{(k^2+k'^2+2)(k'^2+1)^2}dk'\nonumber\\
&=&\frac{16}{\pi}\frac{k}{(k^2+1)^2}\int\limits_{0}^{\infty}\left[\frac{d^2}{dk'^2}+\frac12\right]
\left[\frac{k'}{(k^2+k'^2+2)(k'^2+1)^2}\right]\delta(k'-k)\,dk'
=\frac{4}{\pi}\frac{k^2(k^4+40k^2-29)}{(k^2+1)^7}.\label{eq:A2k}
\end{eqnarray}
Here, we use twice the integration by parts and the fact that the integrand
vanishes at the ends of interval. Thus, the final expression for
$A(k)=A^{I}(k)+A^{II}(k)$ results from the addition of Eq.~(\ref{eq:A1k}) and
Eq.~(\ref{eq:A2k}):
\begin{equation}
A(k)=\frac{8}{\pi}\frac{k^2(6k^2+5)}{(k^2+1)^7}.
\end{equation}
Computation of $B(k)$ is slightly more involved since the resulting function
appears to be non-analytic:
\begin{eqnarray}
B(k)&=&-\frac12\int\limits_{0}^{\infty} 
\frac{\langle\phi_0|x^2|\phi_+(k)\rangle\langle\phi_+(k)|x|\phi_-(k')\rangle\langle\phi_-(k')|x|\phi_0\rangle}
{2\omega_0(k')-\omega_0(k)}dk'\nonumber\\
&=&-\frac{32}{\pi}\frac{k^2}{(1+k^2)^3}\int\limits_{0}^{\infty} 
\frac{1}{2k'^2-k^2+1}\frac{k'}{(k'^2+1)^2}
\left[\frac{d\delta(k'-k)}{dk'}-\frac{1}{\pi}\frac{4k'}{(k'^2-k^2)^2}\right]dk'\nonumber\\
&=&-\frac{32}{\pi}\frac{k^2}{(1+k^2)^3}\left[\lambda_3(k)-\frac{4}{\pi}\lambda_4(k)\right]
=-\frac{256}{\pi}\frac{k^2}{(1+k^2)^7}\left\{
\begin{array}{ll}
\sqrt{1-k^2}\left[\sqrt2-\sqrt{1-k^2}\right],&k<1,\\
k^2-1&k>1.
\end{array}
\right.
\end{eqnarray}
Before the evaluation of $\chi_{qee}^{(\omega,\omega)}$ it is instructive to
analyze the properties of the spectral functions $A^{I}(k)$, $A^{II}(k)$, and
$B(k)$. They describe different excitation pathways (Fig.~\ref{fig:ABk}) in
the system. A peak slightly below $k=\frac12$ related to the existence of the
excitation threshold is common for them.  It is interesting to observe that
$A^{I}(k)$ and $A^{II}(k)$ decay as $k^{-8}$ for large wave-number $k$. There
is, however, a cancellation of two terms in their sum leading to the same
asymptotic behavior as $B(k)$ ($k^{-10}$).

 \begin{figure*}
 \includegraphics[width=17cm]{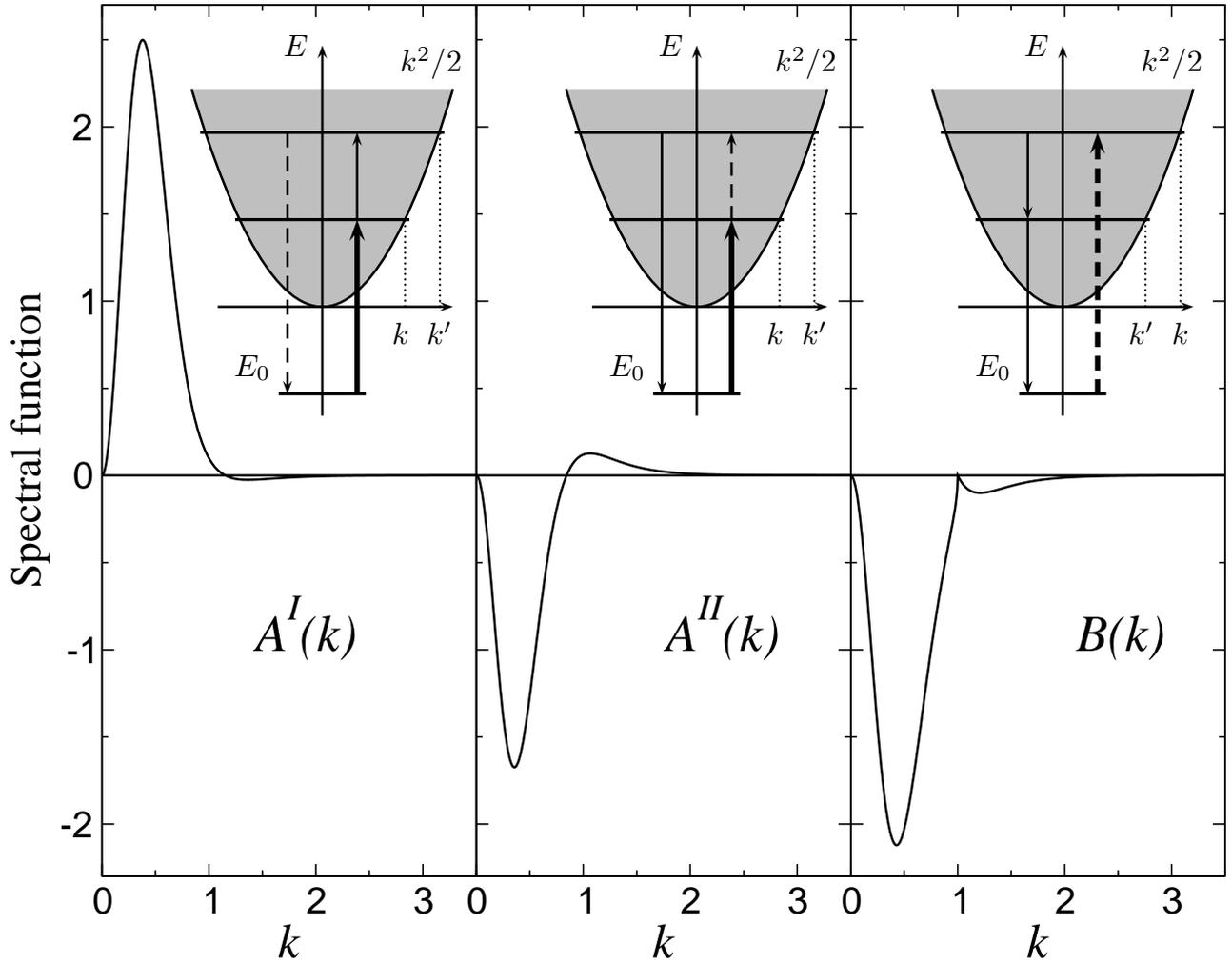}%
 \caption{SHG spectral functions of the one-dimensional hydrogen with a
 $\delta$-function interaction. All quantities are given in atomic
 units. Three panels shows three possible excitation pathways leading to the
 quadrupole polarization of the system. On the insets, thick lines denote
 resonant transitions. Dashed lines correspond to $2\omega$
 transitions. $E_0=-1/2$ denotes the ground state of the system. Unbound
 states have parabolic dispersion.
 \label{fig:ABk}}
 \end{figure*}

We compute the SHG response using the following representation:
\begin{eqnarray}
\chi_{qee}^{(\omega,\omega)}&=&\int\limits_0^\infty dk\frac{A(k)}{\omega-\omega_0(k)+i\Gamma}
+\int\limits_0^\infty dk\frac{-A(k)}{\omega+\omega_0(k)+i\Gamma}
+\int\limits_0^\infty dk\frac{B(k)}{\omega-\omega_0(k)/2+i\Gamma/2}
+\int\limits_0^\infty dk\frac{-B(k)}{\omega+\omega_0(k)/2+i\Gamma/2}\nonumber\\
&=&\frac12\int\limits_{-\infty}^\infty dk\left[\frac{A(k)}{\omega-\omega_0(k)+i\Gamma}
-\frac{A(k)}{\omega+\omega_0(k)+i\Gamma}
+\frac{B(k)}{\omega-\omega_0(k)/2+i\Gamma/2}
- \frac{B(k)}{\omega+\omega_0(k)/2+i\Gamma/2}\right],\label{eq:chi}
\end{eqnarray}
where $\Gamma>0$ is infinitesimally small broadening of the states that ensures
the causality of the response function. We transform the integration to the
whole real axis using the fact that $A(k)$ and $B(k)$ are even functions. In
this form the residue theorem can easily be applied. It is convenient to
introduce following auxiliary functions:
\begin{eqnarray}
a(\omega)&=&\frac12\int\limits_{-\infty}^\infty
dk\frac{A(k)}{\omega-\omega_0(k)+i\Gamma}
=\frac4\pi\int\limits_{-\infty}^\infty dk\,\frac{k^2(6k^2+5)}
{(k^2+1)^7(\omega-1/2(k^2+1)+i\Gamma)},\label{eq:aw}\\
b(\omega)&=&\frac12\int\limits_{-\infty}^\infty
dk\frac{B(k)}{\omega-\omega_0(k)/2+i\Gamma/2}=b_1(\omega)+b_2(\omega),\label{eq:bw}
\end{eqnarray}
and to split the integration of $B(k)$ into two parts:
\begin{subequations}
\begin{eqnarray}
b_1(\omega)&=&-\frac{16}{\pi}\int\limits_{-\infty}^\infty
dk\frac{k^2\lambda_3(k)}{(1+k^2)^3(\omega-\omega_0(k)/2+i\Gamma/2)}
=-\frac{16}{\pi}\int\limits_{-\infty}^\infty
dk\frac{k^2(7k^2-1)}{(1+k^2)^7(\omega-1/4(k^2+1)+i\Gamma/2)},\label{eq:b1w}\\
b_2(\omega)&=&\frac{64}{\pi^2}\int\limits_{-\infty}^\infty
dk\frac{k^2}{(1+k^2)^3(\omega-\omega_0(k)/2+i\Gamma/2)}\lambda_4(k).\label{eq:b2w}
\end{eqnarray}
\end{subequations}

 \begin{figure*}
 \includegraphics[width=17cm]{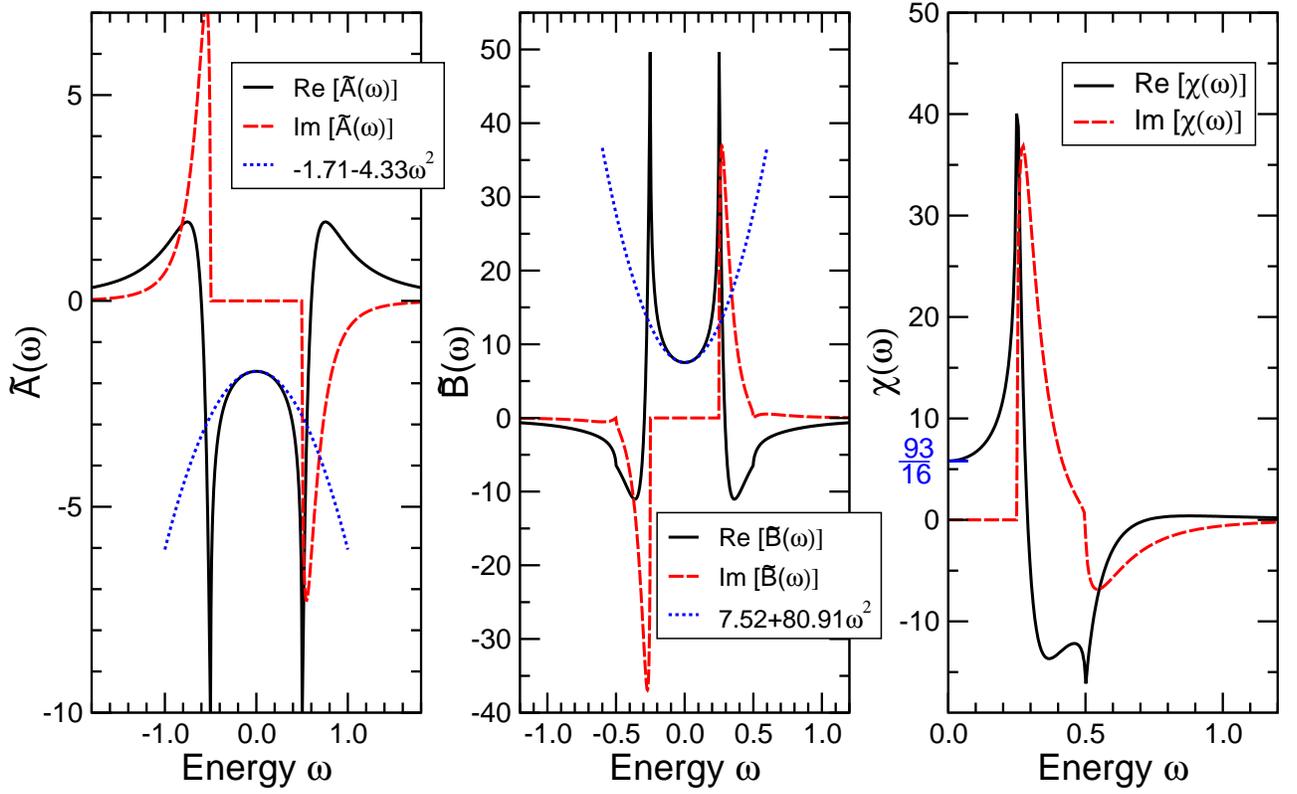}%
 \caption{SHG response functions of the one-dimensional hydrogen with a
 $\delta$-function interaction. The axes are labeled in atomic units. Dotted
 curves represent parabolic dispersion of the response functions in the
 vicinity of $\omega=0$, as given by Eqs.~(\ref{eq:A-disp},\ref{eq:B-disp}).
 \label{fig:ABw}}
 \end{figure*}

Thus, the response function can be written as
$\chi_{qee}^{(\omega,\omega)}=\tilde{A}(\omega)+\tilde{B}(\omega)$, with
$\tilde{A}(\omega)=a(\omega)+a^*(-\omega)$, and
$\tilde{B}(\omega)=b(\omega)+b^*(-\omega)$.  Although one can apply
Sokhotsky's formula in order to quickly get its imaginary part, we will
explicitly perform the integrations in order to obtain the real part. The
calculations are quite cumbersome involving numerous applications of the
residue theorem. Since $\lambda_4(k)$ is not an analytic function, we were
forced to use its integral representation and to exchange the order of
integrations. Details of these calculations are presented in
Appendix~\ref{sec:apb}. Thus, final expressions are given by:
\begin{subequations}
\label{eq:Aw}
\begin{eqnarray}
\mathrm{Im}\tilde{A}(\omega)&=&\frac{1}{16\omega^7}\left\{
\begin{array}{cr}
(1-12|\omega|)\sqrt{2|\omega|-1}&|\omega|>\frac12,\\
0&|\omega|<\frac12;\\
\end{array}
\right.\\
\mathrm{Re}\tilde{A}(\omega)&=&\frac{53\omega^4+44\omega^2-104}{64\omega^6}+\frac{1}{16\omega^7}\left\{
\begin{array}{cr}
-\sqrt{1-2\omega}(1-12\omega)&\omega<-\frac12,\\
\sqrt{1+2\omega}(1+12\omega)-\sqrt{1-2\omega}(1-12\omega)&|\omega|<\frac12,\\
\sqrt{1+2\omega}(1+12\omega)&\omega>\frac12;\\
\end{array}
\right.
\end{eqnarray}
\end{subequations}
\begin{subequations}
\label{eq:Bw}
\begin{eqnarray}
\mathrm{Im}\tilde{B}(\omega)&=&\frac{\sqrt{4|\omega|-1}}{16\omega^7}\left\{
\begin{array}{cr}
2|\omega|-1&|\omega|>\frac12,\\
2|\omega|-1+\sqrt{1-2|\omega|}&\frac12>|\omega|>\frac14,\\
0&|\omega|<\frac14;\\
\end{array}
\right.\\
\mathrm{Re}\tilde{B}(\omega)&=&-\frac{53\omega^4+12\omega^2-8}{64\omega^6}\nonumber\\
&&-\frac{1}{16\omega^7}\left\{
\begin{array}{cr}
\sqrt{1-4\omega}\left[(2\omega-1)+\sqrt{1-2\omega}\right]+\sqrt{(2\omega+1)(4\omega+1)}&-\omega>\frac12,\\
\sqrt{1-4\omega}\left[(2\omega-1)+\sqrt{1-2\omega}\right]&\frac12>-\omega>\frac14,\\
\sqrt{1-4\omega}\left[2\omega-1+\sqrt{1-2\omega}\right]
+\sqrt{1+4\omega}\left[2\omega+1-\sqrt{1+2\omega}\right]&|\omega|<\frac14,\\
\sqrt{1+4\omega}\left[(2\omega+1)-\sqrt{1+2\omega}\right]&\frac14<\omega<\frac12,\\
\sqrt{1+4\omega}\left[(2\omega+1)-\sqrt{1+2\omega}\right]-\sqrt{(2\omega-1)(4\omega-1)}&\omega>\frac12.\\
\end{array}
\right.
\end{eqnarray}
\end{subequations}
\end{widetext} 

\section{Low and high frequency limits\label{sec:limits}}
Qualitatively, the properties of obtained response function
(Fig.~\ref{fig:ABw}) can be understood from general principles. As expected,
it contains both real and imaginary parts related by the Kramers-Kr\"{o}nig
relations, which are well known to hold also for the second harmonic
response~\cite{Scandolo1995}. The real part of the function is associated with
the conversion between the fundamental and second-harmonic frequencies, while
its imaginary part is associated with the removal of energy from the
electromagnetic field~\cite{Nastos2006}. In our case, the Kramers-Kr\"{o}nig
relations directly follow from Eq.~(\ref{eq:chi}). No optical transitions are
possible for the energy of photon below the gap between the ground and unbound
excited states. Therefore, no energy absorption is possible in this
regime. $\mathrm{Im} \tilde{A}(\omega)$ vanishes for $\omega<1/2$ and
$\mathrm{Im} \tilde{B}(\omega)$ vanishes for $\omega<1/4$. $\mathrm{Re}
\tilde{A}$ is sharply peaked at the edge of the continuum: $\mathrm{Re}
\tilde{A}(\frac12)=-9.33$, its frequency derivative has an infinite
discontinuity $\mathrm{Re} \tilde{A}'(\omega=\frac12-)=-\infty$, $\mathrm{Re}
\tilde{A}'(\omega=\frac12+)=213.37$.  Similar behavior is found for
$\mathrm{Re} \tilde{B}$ at $\omega=\frac14$: $\mathrm{Re}
\tilde{B}(\frac14)=52.14$, $\mathrm{Re} \tilde{B}'(\omega=\frac14-)=\infty$,
$\mathrm{Re} \tilde{B}'(\omega=\frac14+)=-2365.37$. Similarly to the imaginary
part, the real part of the response function remains finite and continuous for
all values of $\omega$ and has a well defined static limit:
\begin{eqnarray}
\mathrm{Re}\tilde{A}(\omega)&=&-\frac{219}{128}-\frac{4433}{1024}\omega^2+\ldots\label{eq:A-disp}\\
\mathrm{Re}\tilde{B}(\omega)&=&\frac{963}{128}+\frac{82849}{1024}\omega^2+\ldots\label{eq:B-disp}\\
\mathrm{Re}\chi_{qee}^{(\omega,\omega)}&=&\frac{93}{16}+\frac{4901}{64}\omega^2+\ldots
\end{eqnarray}
At small $\omega$, one observes a quadratic dependence of the SHG response on
the frequency of incident photon. This behavior was already found in many
realistic systems and proved analytically~\cite{Bishop1991,Bishop1996}.

Let us now obtain the static limit closely following the derivation of
\cite{Fernandez1985}. We consider the one-dimensional stationary
Schr\"{o}dinger equation:
\begin{equation}
H\psi=\epsilon\psi,\hspace{1em}
H=-D^2+V(x),\hspace{1em}
D=d/dx,\label{eq:Schr-II}
\end{equation}
where the boundary conditions are supposed to be 
\[
\psi(x\rightarrow\pm\infty)=0,
\]
and $V(x)=-\delta(x)+\lambda x$.  Using the method of Swenson and
Danforth~\cite{Swenson1972} one can obtain recursive equations
\begin{eqnarray}
\lefteqn{X_0^{(N-1)}=(N-1)(N-2)X_0^{(N-3)},}&&\label{eq:Xn}\\
\lefteqn{X_p^{(N-1)}=(N-1)(N-2)X^{(N-3)}_p}\nonumber\\
&&+4\sum_{q=1}^{p}\frac{X^{(1)}_{q-1}X^{(N-1)}_{p-q}}{q}-\frac{2(2N+1)}{N}X^{(N)}_{p-1}.
\label{eq:Xpn}
\end{eqnarray}
for the expectation values of of $x^N$, i.e.  $X^{(N)}=\langle x^N \rangle$ at
different orders of the Rayleigh-Schr\"{o}\-din\-ger perturbation theory:
\begin{equation}
\epsilon=\sum_{p=0}^\infty \epsilon_p\lambda^p,\hspace{1em}
\epsilon_0=-\frac14,\hspace{1em}
X^{(N)}=\sum_{p=0}^\infty X^{(N)}_p\lambda^p.\label{eq:expan}
\end{equation}
From $X^{(0)}_0=1$ and $X^{(1)}_0=0$ one obtains with a help of Eq.~(\ref{eq:Xn})
\[
X^{(2)}_0=2,\hspace{1em}
X^{(4)}_0=24,\hspace{1em}
X^{(6)}_0=720,\ldots,\hspace{1em}
X^{(2N+1)}_0=0.
\]
Starting from these values other coefficients can be recursively computed
using Eq.~(\ref{eq:Xpn}). Simple symmetry consideration show that
$X^{(N)}_p=0$ if $N+p$ is an odd number. First few nonzero values are 
\begin{eqnarray*}
X^{(1)}_1=-10,\hspace{1em}X^{(3)}_1=-168,\\
X^{(2)}_2=744,\hspace{1em}X^{(2)}_4=36960,\\
X^{(3)}_1=-3520,\hspace{1em}X^{(3)}_3=-184080.
\end{eqnarray*}
The knowledge of $X^{(N)}_p$ allows to compute the static response functions.
Let us make a rescaling $q=x/2$, $\phi(q)=\psi(x)$, $F=-4\lambda$,
$E=2\epsilon$, and $E_0=2\epsilon_0=-\frac12$ in order to bring
Eq.~(\ref{eq:Schr-II}) to the form:
\begin{equation}
H_0\phi(q)=-\phi''/2-\delta(q)\phi(q)-Fq\phi(q)=E\phi(q).
\end{equation}
As a consequence an expansion 
\begin{equation}
Q^{(N)}=\sum_p^\infty Q^{(N)}_pF^p,
\end{equation}
must be compared with Eq.~(\ref{eq:expan}) yielding:
\begin{equation}
Q^{(N)}_p=(-1)^p\frac{X^{(N)}_p}{2^N4^p}.
\end{equation}
Hence, the first few values are:
\begin{eqnarray*}
\alpha(0)&=&Q^{(1)}_1=-\frac{1}{8}X^{(1)}_1=\frac54,\\
2\chi_{qee}^{(0,0)}&=&Q^{(2)}_2=\frac{1}{2^24^2}X^{(2)}_2=\frac{93}{8}.
\end{eqnarray*}
We see that $Q^{(1)}_1$ correctly reproduces the static polarizability of our
model system. The frequency dependent polarizability was obtained
in~\cite{Postma1984,Maize2004}. Taking a limit $\omega\rightarrow0$ yields
$\alpha(0)=\frac54$. There is an additional factor of 2 that must be
multiplied with $\chi_{qee}^{(0,0)}$ in order to equate it to
$Q^{(2)}_2$. Appearance of this factor is due to different representations of
electric fields for static and dynamic case used in \cite{Orr}. The resulting
difference in terminology is discussed in details in \cite{Shen-Sec},
alternatively one can see it from the difference in factor
$K(\omega_1,\omega_2)$ adopted by Orr and Ward for static and dynamic cases
(for SHG $K(\omega,\omega)=1/2$, while in the static case $K(0,0)=1$, see
Tab.~1 of \cite{Orr}).

Following the prescription of~\cite{Scandolo1995} the $\omega\rightarrow
\infty$ behavior will be obtained by 
i) defining the $\chi_{qee}^{(\omega,\omega)}$ in terms of its Fourier
transform:
\begin{equation}
\chi_{qee}^{(\omega,\omega)}
=\int d\tau^+\int d\tau^-G^{(2)}(t_1,t_2)e^{i\omega\tau^+},
\label{eq:fourier}
\end{equation}
where $\tau^+=t_1+t_2$, $\tau^-=(t_1-t_2)/2$ and 
\begin{eqnarray}
G^{(2)}(t_1,t_2)&\propto& -f(t_1,t_2)g(t_1,t_2),\\
f(t_1,t_2)&=&\theta(t_1)\theta(t_2-t_1)+\theta(t_2)\theta(t_1-t_2)
\end{eqnarray}
ii) integrating Eq.~(\ref{eq:fourier}) by parts
\begin{equation}
\chi_{qee}^{(\omega,\omega)}=-\sum_m\frac{[\int d\tau^-\frac{\partial^m}
{\partial\tau^{+m}}G^{(2)}(t_1,t_2)]_{\tau^+\rightarrow0^+}}
{(-i\omega)^{m+1}}\label{eq:chi-asympt}
\end{equation}
iii) seeking the lowest nonvanishing order of the time derivative
$\frac{\partial^n}{\partial {\tau^+}^n} g(t_1,t_2)$ of the correlation
function
\begin{equation}
g(t_1,t_2)=\langle\left[x(-t_2),[x(-t_1),x^2]\right]\rangle_0.
\end{equation}
Since we consider here the one-dimensional case, spatial indices are omitted.
After some algebra one can show that the first nonzero term is:
\begin{equation}
\frac{\partial^2}{\partial {\tau^+}^2} g(t_1,t_2)=2\frac{\partial^2 g}{\partial
    t_1\partial t_2}=-4.
\end{equation}
Thus, the first nonvanishing derivative in Eq.~(\ref{eq:chi-asympt}) is of the
third order
\[
\frac{\partial^3}{\partial \tau^{+3}}G^{(2)}(t_1,t_2)\propto
4[\delta(t_1)\theta(t_2-t_1)+\delta(t_2)\theta(t_1-t_2)]
\]
By inserting this expression in Eq.~(\ref{eq:chi-asympt}), performing the
integration and taking the limit we obtain 
\[
\chi_{qee}^{(\omega,\omega)}\propto\frac{1}{\omega^4},
\]
for $\omega\rightarrow\infty$. The same behavior is seen indeed from
Eqs.~(\ref{eq:Aw},\ref{eq:Bw}). One can also note a delicate cancellation of
$\omega^{-2}$ terms in the sum $\bar{A}(\omega)+\bar{B}(\omega)$.

  Another important conclusion that immediately follows from our asymptotic
analysis is the prefactor of $1/\omega^4$ decay. In the SHG case due to dipole
transitions the response function $\chi_{i,jk}^{(\omega,\omega)}$ behaves as
$\langle\frac{\partial^3{V}}{\partial x_i\partial x_j\partial
x_k}\rangle_0\frac{1}{\omega^6}$, where the average is performed in the ground
state of the system~\cite{Scandolo1995}. The potential $V$ describes the
electron-ion interaction. For the third harmonic generation one obtains in the
same way
$\chi_{i,jkl}^{(\omega,\omega,\omega)}=\langle\frac{\partial^4{V}}{\partial
x_i\partial x_j\partial x_k\partial
x_l}\rangle_0\frac{1}{\omega^8}$,~\cite{Rapapa1996}. It is clear that in both
cases the potential and its partial derivatives are system specific. In
contrast, the quadrupole SHG response function does not contain derivatives of
the potential, it only depends on the system density in the high frequency
limit.

\section{Conclusions}
In this work, we analyzed the singular structure of the second-order response
function based on the sum-over-states expression. As an illustration of the
method the non-local second harmonic generation is obtained for the
one-dimensional hydrogen atom with a $\delta$-function interaction. It
provides a new paradigm where the sum over continuum states can be
analytically evaluated.  Finally we analyze the static and high-frequency
limits, which we are also able to compute using different methods.

Our results lead to the fast algorithm for the calculation of non-linear
response using the sum-over-state approach. The analytic expression for the
SHG response can be used as a valuable test of the numerical convergence of
SOS methods. The analysis of the high-frequency behavior shows a universal
$\omega^{-4}$ behavior, which can easily be tested experimentally.

\section{Acknowledgments}
The author wishes to thank Prof. R. Broer and Dr. P. L. de Boeij for very fruitful
discussions.  The author is indebted to Dr. Y. Pavlyukh for his numerous suggestions on the
mathematical methods and for a critical reading of the manuscript.

  \appendix
\begin{widetext}
\section{Matrix elements\label{sec:apa}}
 One of the simplest matrix elements is the bound-free transition dipole
 moment. The integrand belongs to the space of square-integrable
 functions. The integration can be done using elementary methods:
\begin{equation}
\langle\phi_0|x|\phi_-(k)\rangle=\frac1{\sqrt{\pi}}\int_{-\infty}^{\infty}dx e^{-|x|}x\sin(kx)
=-\frac2{\sqrt{\pi}}\frac{d}{dk}\int_{0}^{\infty}dx
e^{-x}\cos(kx)=\frac4{\sqrt{\pi}}\frac{k}{(k^2+1)^2}.
\label{eq:me-d0m}
\end{equation}
Similarly, one obtains the matrix element for the quadrupole transition between
the ground and even excited state:
\begin{eqnarray}
\langle\phi_0|x^2|\phi_+(k)\rangle&=&\frac1{\sqrt{\pi}}\frac{1}{\sqrt{1+k^2}}
\int_{-\infty}^{\infty}dx e^{-|x|}x^2\left(\cos(kx)-\frac1{k}\sin(k|x|)\right)
\nonumber\\
&=&-\frac2{\sqrt{\pi}}\frac{1}{\sqrt{1+k^2}}\left(
\frac{d^2}{dk^2}\int_{0}^{\infty}dx
e^{-x}\cos(kx)-\frac1{k}\frac{d^2}{dk^2}\int_{0}^{\infty}dx
e^{-x}\sin(kx)\right)\nonumber\\
&=&-\frac8{\sqrt{\pi}}\frac{1}{\sqrt{1+k^2}}\frac{1}{(1+k^2)^2}.\label{eq:me-q0p}
\end{eqnarray}
All other matrix elements mentioned in this manuscript cannot be represented
in terms of only elementary functions. This is due to the fact that free
particle states are not square integrable, but rather normalized on the
$\delta$-function. Thus, the quadrupole transition moment between the odd
states can be computed as follows:
\begin{eqnarray}
\langle\phi_-(k')|x^2|\phi_-(k)\rangle&=&\frac1{\pi}\int\limits_{-\infty}^{\infty} dx\,\sin(k'x)x^2\sin(kx)
=-\frac{1}{\pi}\frac{1}{(2i)^2}\frac{d^2}{{dk^2}}\int\limits_{-\infty}^{\infty}
dx\,\left(e^{ik'x}-e^{-ik'x}\right)\left(e^{ikx}-e^{-ikx}\right)\nonumber\\
&=&\frac{d^2}{{dk^2}}\left[\delta(k+k')-\delta(k'-k)\right]=-\delta^{(2)}(k'-k),\label{eq:me-qmm}
\end{eqnarray}
where the last transition is valid because of $k,\,k'>0$ in our case.  The
calculation of the dipole transition moments between the free states of
different parity can be done as follows:
\begin{eqnarray}
\langle\phi_+(k)|x|\phi_-(k')\rangle&=&\frac{1}{\pi}\frac{k}{\sqrt{1+k^2}}
\left[\int\limits_{-\infty}^{\infty}dx\,\cos(kx)x\sin(k'x)-\frac1{k}
\int\limits_{-\infty}^{\infty}dx\,\sin(k|x|)x\sin(k'x)\right]\nonumber\\
&=&-\frac{k}{\sqrt{1+k^2}}\left[\delta^{(1)}(k'-k)-\frac4{\pi}\frac{k'}{(k'^2-k^2)^2}\right],
\label{eq:me-dpm}
\end{eqnarray}
using following integrals
\begin{eqnarray*}
\frac1{\pi}\int\limits_{-\infty}^{\infty}dx\,\cos(kx)x\sin(k'x)
&=&-\frac1{\pi}\frac1{2^2}\frac{d}{dk'}\int\limits_{-\infty}^{\infty}
dx\,\left(e^{ik'x}+e^{-ik'x}\right)\left(e^{ikx}+e^{-ikx}\right)\\
&=&-\frac{d}{dk'}\left[\delta(k+k')+\delta(k'-k)\right]=-\delta^{(1)}(k'-k),\mbox{for
  $k,\,k'>0$}\\
\frac1{k}\int\limits_{-\infty}^{\infty}dx\,\sin(k|x|)x\sin(k'x)
&=&-\frac1{k}\frac{d}{dk'}\int\limits_{-\infty}^{\infty}dx\,\mathrm{sgn}(x)\sin(kx)\cos(k'x)\\
&=&-\frac1{k}\frac{1}{4i}\frac{d}{dk'}\int\limits_{-\infty}^{\infty}dx\,\mathrm{sgn}(x)
\left(e^{ik'x}+e^{-ik'x}\right)\left(e^{ikx}-e^{-ikx}\right)\\
&=&-\frac1{k}\frac{d}{dk'}\left[\frac1{k'+k}-\frac1{k'-k}\right]
=\frac{d}{dk'}\frac{2}{k'^2-k^2}=-\frac{4k'}{(k'^2-k^2)^2},
\end{eqnarray*}
where we used a property
$\int_{-\infty}^{\infty}dx\,\mathrm{sgn}(x)e^{ikx}=2i/k$. The spectral
functions $A^{I}(k)$, $A^{II}(k)$, and $B(k)$ contain contribution from four
integrals ($\lambda_i(k)$, $i=1\ldots4$). Integrals $\lambda_1(k)$ and
$\lambda_3(k)$ are evaluated by the integration by parts and $\lambda_2(k)$,
$\lambda_4(k)$ are evaluated by the contour integration along the real axis
closed by an infinite semi-circle in the upper half of the complex plane
(Fig.~\ref{fig:l4}), as follows:
\begin{eqnarray}
\lambda_1(k)&=&\int\limits_{0}^{\infty}\frac{1}{2k^2-q^2+1}\frac{q^2}{(1+q^2)^3}
\frac{d\delta(q-k)}{dq}dq
=-\int\limits_{0}^{\infty}\delta(q-k)\frac{d}{dq}
\left[\frac{1}{2k^2-q^2+1}\frac{q^2}{(1+q^2)^3}\right]dq\nonumber\\
&=&\frac{2k(k^2-1)}{(k^2+1)^5},\\
\lambda_2(k)&=&\int\limits_{0}^{\infty}\frac{1}{2k^2-q^2+1}\frac{q^2}{(1+q^2)^3}
\frac{1}{(k^2-q^2)^2} dq
=\frac12\int\limits_{-\infty}^{\infty}\frac{1}{2k^2-q^2+1}\frac{q^2}{(1+q^2)^3}
\frac{1}{(k^2-q^2)^2} dq\nonumber\\
&=&\frac{1}{2}2\pi i\mathrm{Res}_{q=i}\left[\frac{1}{2k^2-q^2+1}\frac{q^2}{(1+q^2)^3}
\frac{1}{(k^2-q^2)^2}\right]=\frac{\pi}{32}\frac{k^4+12k^2-23}{(k^2+1)^5},\\
\lambda_3(k)&=&\int\limits_{0}^{\infty}\frac{1}{2q^2-k^2+1}\frac{q}{(q^2+1)^2}\frac{d\delta(q-k)}{dq}dq
=-\int\limits_{0}^{\infty}\delta(q-k)\frac{d}{dq}\left[\frac{1}{2q^2-k^2+1}\frac{q}{(q^2+1)^2}\right]dq
\nonumber\\
&=&\frac{7k^2-1}{(k^2+1)^4},\\
\lambda_4(k)&=&\int\limits_{0}^{\infty}\frac{1}{2q^2-k^2+1}\frac{q^2}{(q^2+1)^2(q^2-k^2)^2}dq
=\frac12\int\limits_{-\infty}^{\infty}\frac{1}{2q^2-k^2+1}\frac{q^2}{(q^2+1)^2(q^2-k^2)^2}dq\nonumber\\
&=&\frac{1}{2}2\pi i\left\{
\begin{array}{ll}
\mathrm{Res}(k=i)+\mathrm{Res}\left(k=i\sqrt{\frac{1-k^2}{2}}\right),&k<1\\
\mathrm{Res}(k=i),&k>1
\end{array}
\right.
=-\frac{\pi}{4}\frac{k^2-7}{(1+k^2)^4}
-\frac{4\pi}{(1+k^2)^4}
\left\{
\begin{array}{ll}
\sqrt{\frac{1-k^2}{2}},&k<1\\
0,&k>1
\end{array}
\right..
\end{eqnarray}

 \begin{figure*}
 \includegraphics[width=17cm]{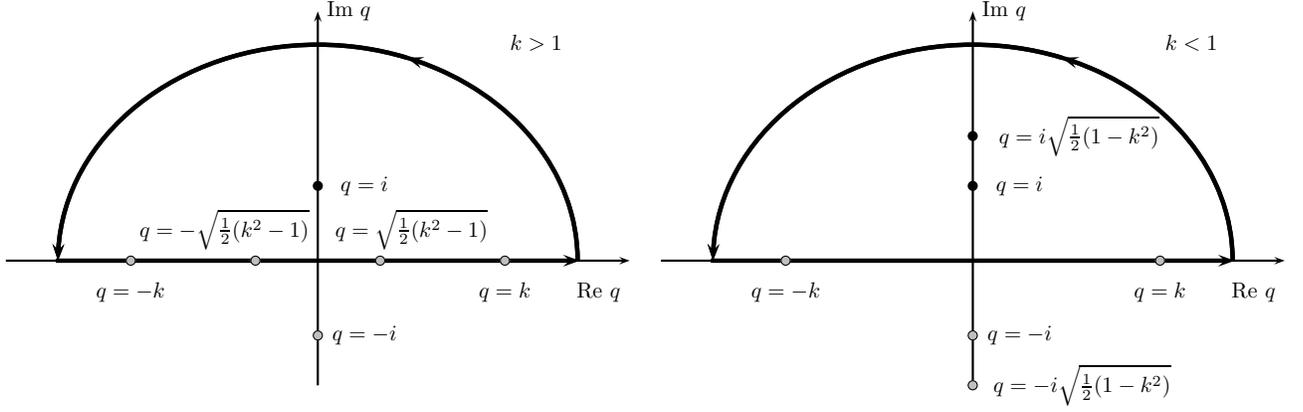}%
 \caption{Integration of $\lambda_4(\omega)$. Poles that contribute to the
 integral are shown in black. $q=\pm i$ and $q=\pm k$ are second order poles.
 $q=\pm\sqrt{\frac12(k^2-1)}$ are simple poles that lay on the imaginary axis
 for $k<1$ and on the real axis for $k>1$. One can verify that poles of each
 pair have residues of opposite sign; therefore, their contributions to the
 integral cancel if the singularities are situated on the real axis.
 \label{fig:l4}}
 \end{figure*}
\end{widetext}
\section{Hilbert transforms\label{sec:apb}}
Below, we evaluate integrals $a(\omega)$, $b_1(\omega)$ and
$b_2(\omega)$. They are defined by
Eqs.~(\ref{eq:aw},\ref{eq:b1w},\ref{eq:b2w}), respectively. These are quite
cumbersome calculations, which involve application of the residue theorem. For
$a(\omega)$, the number of poles in the upper complex half-plane encircled by
the contour of integration depends on the value of frequency $\omega$
(Fig.~\ref{fig:aw}). When $\omega<\frac12$, i.e. the energy of incident photon
is insufficient to generate optical transition from the ground state to the
continuum of unbound states. The imaginary part of the function $a(\omega)$
vanishes. While the real part has a contribution from two poles. For the
energy of an incident photon above the ionization threshold,
i.e. $\omega>\frac12$, one pole contributes to the real and one pole
contributes to the imaginary part of the function.
\begin{widetext}
\begin{eqnarray*}
\mathrm{Re}\,a(\omega)&=&\pi i\left[\left\{
\begin{array}{cl}
\mathrm{Res}_{k=i}&\omega>\frac12,\\
\mathrm{Res}_{k=i}+\mathrm{Res}_{k=i\sqrt{1-2\omega}}&\omega<\frac12
\end{array}
\right.\right]\frac{A(k)}{\omega-\omega_0(k)+i\Gamma}\\
&=&\frac{147\omega^6+106\omega^5+86\omega^4+88\omega^3+184\omega^2-208\omega+16}{256\omega^7}
-\frac{1}{16\omega^7}\left\{
\begin{array}{cl}
0&\omega>\frac12,\\
\sqrt{1-2\omega}(-12\omega+1)&\omega<\frac12;
\end{array}
\right.\\
\mathrm{Im}\,a(\omega)&=&\pi i\left[\left\{
\begin{array}{cl}
\mathrm{Res}_{k=\sqrt{2\omega-1}}&\omega>\frac12,\\
0&\omega<\frac12
\end{array}
\right.\right]\frac{A(k)}{\omega-\omega_0(k)+i\Gamma}
=\frac{1}{16\omega^7}\left\{
\begin{array}{cl}
\sqrt{2\omega-1}(-12\omega+1)&\omega>\frac12,\\
0&\omega<\frac12.
\end{array}
\right.
\end{eqnarray*}
 
 \begin{figure*}
 \includegraphics[width=17cm]{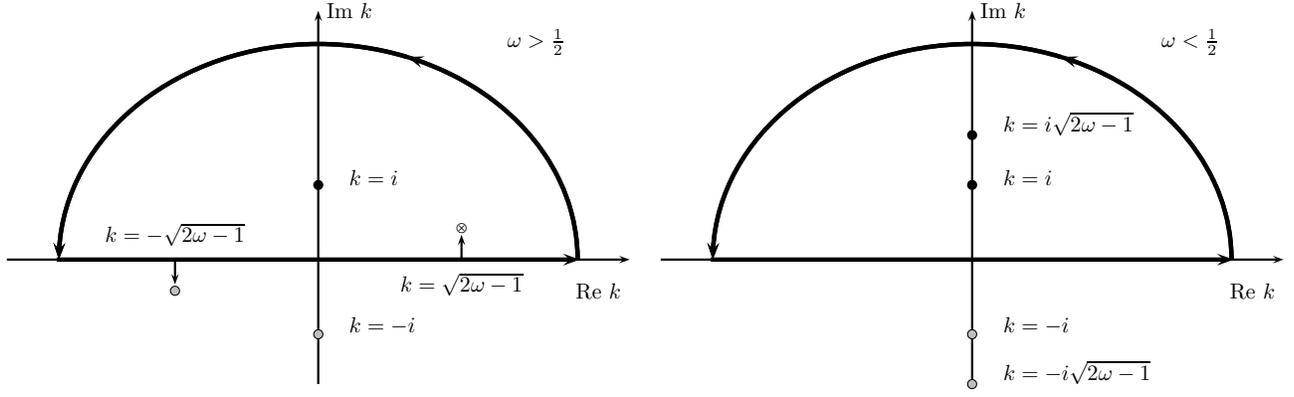}%
 \caption{Integration of $a(\omega)$. $k=\pm i$ are poles of the seventh
 order. $k=\pm\sqrt{2\omega-1}$ are simple poles. It is important to notice
 that due to the presence of small imaginary part
 ($\omega\rightarrow\omega+i\Gamma$, $\Gamma>0$) they are shifted away from
 the real axis even in the case of $\omega>\frac12$. Thus, only one pole of
 this pair is encircled by the contour for any value of $\omega$. However,
 when $\omega>\frac12$, its residue is real, thus contributing to the
 imaginary part of the integral. In all other cases, the poles have imaginary
 residues. This fact is reflected by different notations for singularities
 with real (crossed circle) and imaginary (black circle) residues.
 \label{fig:aw}}
 \end{figure*}

Calculations for $b_1(\omega)$ can be done along the same line with a small
distinction that now $\omega=\frac14$ is a point, in which the function changes
its behavior. This is because $b_1(\omega)$ describes resonant $2\omega$
transitions. The result works out to be:
\begin{eqnarray*}
\mathrm{Re}\,b_1(\omega)&=&
-\frac{28\omega^6+14\omega^5+8\omega^4+6\omega^3+10\omega^2-11\omega+2}{64\omega^7}
-\frac1{64\omega^7}
\left\{
\begin{array}{cl}
0&\omega>\frac14,\\
\sqrt{1-4\omega}&\omega<\frac14;
\end{array}
\right.\\
\mathrm{Im}\,b_1(\omega)&=&\frac1{64\omega^7}
\left\{
\begin{array}{cl}
\sqrt{4\omega-1}(7\omega-2)&\omega>\frac14,\\
0&\omega<\frac14.
\end{array}
\right.
\end{eqnarray*}
The computation of $b_2(\omega)$ brings us to the following double integral:
\[
b_2(\omega)=\frac{32}{\pi^2}\int\limits_{-\infty}^{\infty}dq\,\frac{q^2}{(q^2+1)^2}
\int\limits_{-\infty}^{\infty}dk\,\frac{k^2}{(1+k^2)^3(\omega-\omega_0(k)/2+i\Gamma/2)
(2q^2-k^2+1)(q^2-k^2)^2}.
\]
 \begin{figure*}
 \includegraphics[width=14cm]{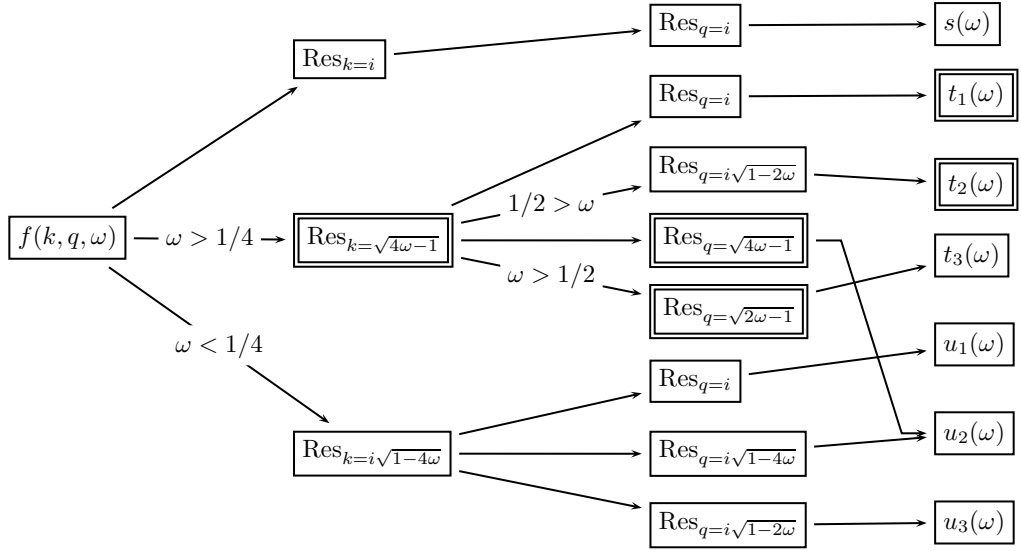}%
 \caption{Scheme that shows how the double integration of $b_2(\omega)$ is
 performed. In both cases, the contour of integration encircles the upper
 complex half-plane. On the first step, we find the residues of poles
 contributing to the integral over $k$. Furthermore, we consider all possible
 ramifications that arise for different values of $\omega$ when integrating
 over $q$. Some branches of the integration tree lead to the same resulting
 function. In order to disentangle contributions to real (imaginary) part of
 the integral, we denote residues that have real (imaginary) value by a single
 (double) frames.
 \label{fig:b2}}
 \end{figure*}
After simple but lengthy consideration of all cases (Fig.~\ref{fig:b2}) we obtain:
\begin{eqnarray*}
\mathrm{Re}\,b_2(\omega)&=&s(\omega)+u_2(\omega)+\left\{
\begin{array}{cl}
t_3(\omega)&\omega>\frac12,\\
0&\frac12>\omega>\frac14,\\
u_1(\omega)+u_3(\omega)&\omega<\frac14;
\end{array}
\right.\\
\mathrm{Im}\,b_2(\omega)&=&\left\{
\begin{array}{cl}
t_1(\omega)&\omega>\frac12,\\
t_1(\omega)+t_2(\omega)&\frac12>\omega>\frac14,\\
0&\omega<\frac14;
\end{array}
\right.
\end{eqnarray*}
where the auxiliary functions ($s(\omega)$, $t_i(\omega)$, and $u_i(\omega)$,
$i=1,\ldots,3$) are given by:
\begin{eqnarray*}
s(\omega)&=&\mathrm{Res}_{q=i}\left(\mathrm{Res}_{k=i}[f(k,q,\omega)]\right)
=-\frac{197\omega^2+50\omega+10}{256\omega^3},\\
t_1(\omega)&=&\mathrm{Res}_{q=i}\left(\mathrm{Res}_{k=\sqrt{4\omega-1}}[f(k,q,\omega)]\right)
=\frac{\sqrt{4\omega-1}(\omega-2)}{64\omega^7},\\
t_2(\omega)&=&
\mathrm{Res}_{q=i\sqrt{1-2\omega}}\left(\mathrm{Res}_{k=\sqrt{4\omega-1}}[f(k,q,\omega)]\right)
=\frac{\sqrt{4\omega-1}\sqrt{1-2\omega}}{16\omega^7},\\
t_3(\omega)&=&
\mathrm{Res}_{q=\sqrt{2\omega-1}}\left(\mathrm{Res}_{k=\sqrt{4\omega-1}}[f(k,q,\omega)]\right)
=\frac{\sqrt{4\omega-1}\sqrt{2\omega-1}}{16\omega^7},\\
u_1(\omega)&=&
\mathrm{Res}_{q=i}\left(\mathrm{Res}_{k=i\sqrt{1-4\omega}}[f(k,q,\omega)]\right)
=-\frac{\sqrt{1-4\omega}(\omega-2)}{64\omega^7},\\
u_2(\omega)&=&
\mathrm{Res}_{q=i\sqrt{1-4\omega}}\left(\mathrm{Res}_{k=i\sqrt{1-4\omega}}[f(k,q,\omega)]\right)
=\mathrm{Res}_{q=\sqrt{4\omega-1}}\left(\mathrm{Res}_{k=\sqrt{4\omega-1}}[f(k,q,\omega)]\right)
=-\frac{7\omega-2}{64\omega^7},\\
u_3(\omega)&=&
\mathrm{Res}_{q=i\sqrt{1-2\omega}}\left(\mathrm{Res}_{k=i\sqrt{1-4\omega}}[f(k,q,\omega)]\right)
=-\frac{\sqrt{1-4\omega}\sqrt{1-2\omega}}{16\omega^7}.
\end{eqnarray*}
Here we denote the function under the integrals as $f(k,q,\omega)$.
\end{widetext} 


\end{document}